\documentclass[11pt,oneside,letterpaper]{article}
\usepackage{graphicx,amssymb,amsmath,anysize,epsfig}

\begin{document}
\title{\bf Comment on ``How the potentials in different gauges yield
the same retarded electric and magnetic fields,"
\\by J. A. Heras [Am. J. Phys. 75, 176 (2007)]}
\author{V. Hnizdo$^{\rm a)}$\\
\it{National Institute for Occupational Safety and Health,
Morgantown, West Virginia 26505} }
\date{}
\maketitle

In a recent paper,$^1$  Heras surveys the Lorenz, Coulomb,
Kirchhoff, velocity, and temporal gauges with a view to explaining
how the potentials of all these gauges yield the same retarded
electromagnetic field, despite the fact that these potentials may
satisfy dynamical equations that do not admit properly retarded
solutions. He claims to show without actually solving the equations
that the potentials satisfy that the ``spurious"
non-causal term generated by the scalar potential of the Coulomb,
Kirchhoff, or velocity gauge is canceled by an equal and opposite
term in the contribution to the electric field that is generated by
the vector potential. With no intention of diminishing the value of
the gauge survey itself, we argue that, given the definitions
of the electric and magnetic fields in terms of electromagnetic
potentials, Heras's cancelation of a non-causal term in the electric
field is an artefact of algebraic manipulation that has no
explanatory content.

Heras  manipulates the equations that, for example, the potentials
$\Phi_C$ and ${\bf A}_C$ of the Coulomb gauge satisfy in four
distinct steps, which in the end  lead to inhomogeneous wave
equations
\begin{eqnarray}
\square^2\left(-{\bf\nabla} \Phi_C-\frac{\partial{\bf A}_C}{\partial
t}\right) &=& \frac{1}{\epsilon_0}{\bf \nabla}\rho
+\mu_0\frac{\partial{\bf J}}{\partial t},\\
\square^2({\bf\nabla \times A}_C) &=& -\mu_0{\bf \nabla\times J}.
\end{eqnarray}
He then writes the retarded solution $-{\bf\nabla}
\Phi_C-\partial{\bf A}_C/\partial t$ of (1) as
\begin{equation}
-\frac{\partial{\bf A}_C}{\partial t}=\frac{1}{4\pi\epsilon_0} \int
\frac{d^3x'}{|{\bf x}-{\bf x}'|}\left[-{\bf
\nabla}'\rho-\frac{1}{c^2}\frac{\partial{\bf J}}{\partial
t'}\right]_{\rm ret} +{\bf\nabla}\Phi_C,
\end{equation}
which must give the electric field ${\bf
E}=-{\bf\nabla}\Phi_C-\partial{\bf A}_C/\partial t$ as
\begin{equation}
{\bf E}=\frac{1}{4\pi\epsilon_0} \int \frac{d^3x'}{|{\bf x}-{\bf
x}'|}\left[-{\bf \nabla}'\rho-\frac{1}{c^2}\frac{\partial{\bf
J}}{\partial t'}\right]_{\rm ret}
\end{equation}
since the instantaneous term $-{\bf\nabla}\Phi_C$ in $-{\bf\nabla}\Phi_C
-\partial{\bf A}_C/\partial t$ is canceled by an
equal and opposite term in $-\partial{\bf A}_C/\partial t$ of (3).
Because of this cancelation, Heras
calls the term $-{\bf\nabla}\Phi_C$ a ``spurious" field, which has
only a ``mathematical", but not ``physical" existence.

We note first that Heras's four steps are not needed to obtain the
wave equations (1) and (2)
--- Maxwell's equations lead directly, without the aid of any
potentials, to inhomogeneous wave equations for the electric and
magnetic fields themselves:$^2$
\begin{eqnarray}
\square^2{\bf E} &=& \frac{1}{\epsilon_0}{\bf \nabla}\rho
+\mu_0\frac{\partial{\bf J}}{\partial t},\\
\square^2{\bf B} &=& -\mu_0{\bf \nabla\times J},
\end{eqnarray}
the retarded solutions of which are
\begin{eqnarray}
{\bf E}&=&\frac{1}{4\pi\epsilon_0} \int\frac{d^3x'}{|{\bf x}-{\bf
x}'|}\left[-{\bf\nabla}'\rho-\frac{1}{c^2}\frac{\partial{\bf
J}}{\partial t'}\right]_{\rm ret},
\\
{\bf B}&=&\frac{\mu_0}{4\pi} \int \frac{d^3x'}{|{\bf x}-{\bf
x}'|}[{\bf \nabla}'{\bf \times J}]_{\rm ret}.
\end{eqnarray}
Equations (1)--(3) then follow immediately on the use in Eqs.\
(5)--(7) of the definitions
\begin{eqnarray}
{\bf E} &=& -{\bf\nabla}\Phi-\frac{\partial{\bf A}}{\partial t},\\
\quad\quad {\bf B} &=& {\bf\nabla \times A},
\end{eqnarray}
where $\Phi$ and $\bf A$ are potentials in an arbitrary gauge. Heras
has to employ the definitions (9) and (10) already when he derives
from Maxwell's equations the dynamical equations for the potentials
of the gauges he considers. The retarded solution (3), written in
the usual manner, thus holds in an arbitrary gauge:
\begin{equation}
-{\bf\nabla}\Phi-\frac{\partial{\bf A}}{\partial
t}=\frac{1}{4\pi\epsilon_0} \int \frac{d^3x'}{|{\bf x}-{\bf
x}'|}\left[-{\bf \nabla}'\rho-\frac{1}{c^2}\frac{\partial{\bf
J}}{\partial t'}\right]_{\rm ret}.
\end{equation}

Equations (7)--(10) ensure  that, irrespective of the gauge used,
the retarded electromagnetic field is given uniquely by its sources
$\rho$ and $\bf J$, but do not help to answer the perhaps
superfluous, but still intriguing, question {\it how} that can
happen when the potentials themselves are solutions of dynamical equations
that do not admit properly retarded solutions. Only constructive
solutions of the dynamical equations for the potentials in terms of
the sources $\rho$ and $\bf J$  can help in this regard, like, for
example, the solution obtained by Jackson$^3$ for the Coulomb-gauge
vector potential ${\bf A}_C$,
\begin{equation}
{\bf A}_C =\frac{\mu_0}{4\pi}\int \frac{d^3x'}{R}\left([{\bf
J}-c\rho\hat{\bf R}]_{\rm ret} +\frac{c^2\hat{\bf
R}}{R}\int_0^{R/c}d\tau \rho({\bf x}',t-\tau)\right),
\end{equation}
where $R=|{\bf x}-{\bf x}'|$ and $\hat{\bf R}=({\bf x}-{\bf x}')/R$.
The partial time derivative of (12) yields
\begin{equation}
-\frac{\partial{\bf A}_C}{\partial t} =\frac{1}{4\pi\epsilon_0}\int
d^3x'\left(\left[\frac{\rho}{R^2}\,\hat{\bf
R}+\frac{1}{cR}\frac{\partial\rho}{\partial t'}\,\hat{\bf
R}-\frac{1}{c^2 R}\,\frac{\partial{\bf J}}{\partial t'}\right]_{\rm
ret}-\frac{\rho}{R^2}\,\hat{\bf R}\right).
\end{equation}
Here, the retarded terms in the integrand give Jefimenko's
expression for the retarded electric field,$^4$ while the last term
yields an instantaneous term ${\bf\nabla}\Phi_C$, which will be
canceled by the equal and opposite term in (9).

In contrast,  it is not difficult to see that the cancelation of the
term $-{\bf\nabla}\Phi_C$ in $\bf E$ by an equal and opposite term
in $-\partial {\bf A}_C/\partial t$ obtained by Heras is merely an
artefact of writing, in Eq.\ (3), the retarded solution
of the inhomogeneous wave equation (1), which is the retarded electric
field $\bf E$ of Eq.\ (4), as $-\partial {\bf A}_C/\partial t
={\bf E}+{\bf\nabla}\Phi_C$.
Using the Lorenz-gauge potentials $\Phi_L$
and ${\bf A}_L$ in Eqs.\ (9) and (11), one could argue {\it \`a la}
Heras that the retarded term $-{\bf\nabla}\Phi_L$ in $\bf E$ is also
a ``spurious" field because it is canceled by an equal and opposite
term in $-\partial{\bf A}_L/\partial t$, the arbitrary exclusion by
Heras of the Lorenz gauge from the four steps of his method
notwithstanding. In general, neither of the
potential-generated terms
$-{\bf\nabla}\Phi$ and $-\partial{\bf A}/\partial t$  alone represents
a physical field, but that does not mean that any of these terms can
be regarded as spurious because both are needed in the
sum $-{\bf\nabla}\Phi-\partial{\bf A}/\partial t$
that is guaranteed  by the mathematical consistency of the
electromagnetic-potential method of solving Maxwell's equations
to yield the physical electric field $\bf E$.
\\

\noindent $^{\rm a)}$Electronic mail: vbh5@cdc.gov. V. Hnizdo has
written this comment in his private capacity. No official support or
endorsement by Centers for Disease Control and Prevention is
intended or should be inferred.
\\
\noindent$^1$J. A. Heras, ``How the potentials in different gauges
yield the same retarded electric and magnetic fields," Am. J. Phys.
{\bf 75}, 176--183 (2007).
\\
\noindent$^2$The inhomogeneous wave equations for $\bf E$ and $\bf
B$ can be obtained by taking curls of Maxwell's curl equations and
decoupling the resulting equations by using the remaining pair of
Maxwell's equations.
\\
\noindent$^3$J. D. Jackson, ``From Lorenz to Coulomb and other
explicit gauge  transformations," Am. J. Phys. {\bf 70}, 917--928
(2002).
\\
\noindent$^4$O. D. Jefimenko, {\it Electricity and Magnetism}
(Electret Scientific, Star City, 1989), 2nd ed.; J. D. Jackson, {\it
Classical Electrodynamics} (Wiley, New York, 1999), 3rd ed., Sec.
6.5.

\end{document}